%% file: panglobal-double-soft.tex
\documentclass[twocolumn,preprintnumbers,superscriptaddress,aps,nofootinbib]{revtex4-2}
\usepackage{amsmath,amssymb,mathtools}
\usepackage{graphicx}
\usepackage{booktabs,cellspace}
\usepackage{color}
\usepackage{hyperref}
\usepackage{xspace}
\usepackage[normalem]{ulem}

\newcommand{\kT}{k_t}

\newcommand{\itilde}{{\tilde \imath}}
\newcommand{\jtilde}{{\tilde \jmath}}

\newcommand{\as}{\ensuremath{\alpha_s}\xspace}
\newcommand{\shower}{\text{shower}}

\newcommand{\TeV}{\ensuremath{\,\mathrm{TeV}}\xspace}

\newcommand{\nc}{\ensuremath{N_\text{\textsc{c}}}\xspace}
\newcommand{\ds}{\text{\textsc{ds}}}
\newcommand{\ps}{\text{\textsc{ps}}}
\newcommand{\dl}{\text{\textsc{dl}}}
\newcommand{\nndl}{\text{\textsc{nndl}}}
\newcommand{\nsl}{\text{\textsc{nsl}}}
\newcommand{\SL}{\text{\textsc{sl}}}
\newcommand{\etabar}{\ensuremath{\bar \eta}\xspace}

\newcommand{\KCMW}{\ensuremath{K_\textsc{cmw}}\xspace}
\newcommand{\PS}{\text{\textsc{(ps)}}}
\newcommand{\rescl}{\ensuremath{r_{\scriptscriptstyle{\text{L}}}}}
\newcommand{\rescg}{\ensuremath{r_{\scriptscriptstyle{\text{G}}}}}

\newcommand{\ourdot}{.}

\newcommand{\order}[1]{\mathcal{O}\left(#1\right)}

\newcommand{\logbook}[2]{}

\definecolor{darkgreen}{rgb}{0,0.4,0}
\definecolor{grey}{rgb}{0.5,0.5,0.5}
\definecolor{orange}{rgb}{0.9,0.5,0.0}
\definecolor{lightblue}{rgb}{0.0,0.5,1.0}
\usepackage[dvipsnames]{xcolor}

\newcommand{\OXaff}{Rudolf Peierls Centre for Theoretical Physics,
  Clarendon Laboratory, Parks Road, Oxford OX1 3PU, UK} 
\newcommand{\CERNaff}{CERN, Theoretical Physics Department, CH-1211 Geneva 23, Switzerland}
\newcommand{\ASCaff}{All Souls College, Oxford OX1 4AL, UK}
\newcommand{\UCLaff}{Department of Physics and Astronomy, University College London, London, WC1E 6BT, UK}
\newcommand{\IPhTAff}{IPhT, Universit\'{e} Paris-Saclay, CNRS UMR 3681,
  CEA Saclay, F-91191 Gif-sur-Yvette, France}

\begin{document}

\title{Parton showering with higher-logarithmic accuracy for soft emissions}

\preprint{CERN-TH-2023-127, OUTP-23-07P}

\author{Silvia Ferrario Ravasio}  \affiliation{\CERNaff}  %
\author{Keith Hamilton}           \affiliation{\UCLaff}  %
\author{Alexander Karlberg}       \affiliation{\CERNaff}  %
\author{Gavin P.\ Salam}          \affiliation{\OXaff}\affiliation{\ASCaff}%
\author{Ludovic Scyboz}           \affiliation{\OXaff}  %
\author{Gregory Soyez} \affiliation{\CERNaff}\affiliation{\IPhTAff}  %

\begin{abstract}
  The accuracy of parton-shower simulations is often a limiting factor
  in the interpretation of data from high-energy colliders.
  We present the first formulation of parton showers with accuracy one
  order beyond state-of-the-art next-to-leading logarithms, for
  classes of observable that are dominantly sensitive to low-energy
  (soft) emissions, specifically non-global observables and subjet
  multiplicities.
  This represents a major step towards general next-to-next-to-leading
  logarithmic accuracy for parton showers.
\end{abstract}

% \pacs{12.38.-t}
\maketitle

Parton showers simulate the repeated branching of quarks and gluons
(partons) from a high momentum scale down to the non-perturbative
scale of Quantum Chromodynamics (QCD).
They are one of the core components of the general-purpose Monte Carlo
event-simulation programs  that are used in almost every experimental and
phenomenological study involving high-energy particle colliders,
such as CERN's Large Hadron Collider (LHC).
Parton-shower accuracy is critical at colliders, both because it
limits the interpretation of data and because of the increasing
importance of showers in training powerful machine-learning based
data-analysis methods.

In the past few years it has become clear that it is instructive to
relate the question of parton-shower accuracy to a shower's ability to
reproduce results from the field of resummation, which sums dominant
(logarithmically enhanced) terms in perturbation theory to all orders
in the strong coupling, $\as$.
Given a logarithm $L$ of some large ratio of momentum scales,
resummation accounts for terms $\as^n L^{n+1-p}$, N$^p$LL in a
leading-logarithmic counting for $L \sim 1/\as$, or $\as^n L^{2n-p}$,
N$^p$DL in a double-logarithmic counting, for $L \sim 1/\sqrt{\as}$.

Several groups have recently proposed parton showers designed to
achieve next-to-leading logarithmic (NLL) and next-to-double logarithmic
(NDL) accuracy for varying sets of
observables~\cite{Bewick:2019rbu,Dasgupta:2020fwr,Forshaw:2020wrq,Nagy:2020rmk,Nagy:2020dvz,Bewick:2021nhc,vanBeekveld:2022zhl,vanBeekveld:2022ukn,Herren:2022jej,vanBeekveld:2023lfu}.
A core underlying requirement is the condition that a shower should
accurately reproduce the tree-level matrix elements for configurations
with any number of low-energy (``soft'') and/or collinear particles,
as long as these particles are well separated in logarithmic phase
space~\cite{Andersson:1988gp,Dreyer:2018nbf,Dasgupta:2020fwr}.
  
In this letter we shall demonstrate a first major
step towards the next order in resummation in a full parton shower,
concentrating on the sector of phase space involving soft partons.
This sector is connected with two important aspects of LHC
simulations, namely the total number of particles produced, and the
presence of soft QCD radiation around leptons and photons
(``isolation''), which is critical in their experimental
identification in a wide range of LHC analyses.
The corresponding areas of resummation theory, for subjet
multiplicity~\cite{Catani:1991hj,Catani:1991pm,Catani:1992tm} and
so-called non-global logarithms~\cite{Dasgupta:2001sh,Dasgupta:2002bw,Banfi:2002hw,Weigert:2003mm,Hatta:2008st,Caron-Huot:2015bja,Forshaw:2009fz,DuranDelgado:2011tp,Schwartz:2014wha,Becher:2015hka,Larkoski:2015zka,Becher:2016mmh,Becher:2016omr,Neill:2016stq,Caron-Huot:2016tzz,Larkoski:2016zzc,Becher:2017nof,AngelesMartinez:2018cfz,Balsiger:2018ezi,Neill:2018yet,Balsiger:2019tne,Balsiger:2020ogy,Hatta:2013iba,Hagiwara:2015bia,Hatta:2020wre,Forshaw:2006fk,Forshaw:2008cq}, have seen
extensive recent developments towards higher accuracy in their own
right, with several groups working either on next-to-next-to-double
logarithmic (NNDL) accuracy, $\as^n L^{2n-2}$, for
multiplicity~\cite{Medves:2022ccw,Medves:2022uii} or next-to-single
logarithmic (NSL) accuracy, $\as^n L^{n-1}$, for non-global
logarithms~\cite{Banfi:2021owj,Banfi:2021xzn,Becher:2021urs,Becher:2023vrh}. 

To achieve NSL/NNDL accuracy for soft-dominated observables, a crucial
new ingredient is that the shower should obtain the correct matrix
element even when there are pairs of soft particles that are
commensurate in energy and in angle with respect to their emitter.
Several groups have worked on incorporating higher-order
soft/collinear matrix elements into parton
showers~\cite{Jadach:2011kc,Hartgring:2013jma,Jadach:2013dfd,Jadach:2016zgk,Li:2016yez,Hoche:2017iem,Hoche:2017hno,Dulat:2018vuy,Campbell:2021svd,Gellersen:2021eci}. 
%
% Jadach:2011kc
% - 1102.5083 - Two real parton contributions to non-singlet kernels for exclusive QCD DGLAP evolution
% Hartgring:2013jma
% - 1303.4974 - Antenna Showers with One-Loop Matrix Elements
% Jadach:2013dfd
% - 1310.6090 - NLO corrections in the initial-state parton shower <<MC>>
% Jadach:2016zgk
% - 1606.01238 - << Dependence of splitting functions on choice of evoln var >>
% Li:2016yez
% - 1611.00013 - A framework for second-order parton showers
% Hoche:2017iem
% - 1705.00742 - << DIRE inclusion of flav changing 1->3, differential >>
% Hoche:2017hno
% - 1705.00982 - << Same as the one above but with NLO 1->2 kernels as well >>
% Dulat:2018vuy
% - 1805.03757 - <<DIRE double-soft paper>>
% Campbell:2021svd
% - 2108.07133 - Towards NNLO+PS matching with sector showers
% Gellersen:2021eci
% - 2110.05964 - Disentangling soft and collinear effects in QCD parton showers
Our approach will be distinct in two respects: firstly, that it is in
the context of a full shower that is already NLL accurate, which is
crucial to ensure that the correctness of any higher-order
matrix element is not broken by recoil effects from subsequent shower
emissions;
and secondly in that we will be able to demonstrate the logarithmic
accuracy for concrete observables through comparisons to known
resummations. 

We will work in the context of the ``PanGlobal'' family of parton
showers, concentrating on the final-state case~\cite{Dasgupta:2020fwr}.
As is common for parton showers, it organises particles into colour
dipoles~\cite{Gustafson:1987rq}, a picture based on the limit of a
large number of colours \nc.
Such showers iterate $2\to 3$ splitting of colour dipoles, each
splitting thus adding one particle to the ensemble, and typically
breaking the original dipole into two dipoles.
The splittings are performed sequentially in some ordering variable,
$v$, for example in decreasing transverse momentum $k_t$.
Given a dipole composed of particles with momenta $\tilde p_i$ and
$\tilde p_j$, the basic kinematic map for producing a new particle $k$
is
\begin{subequations} \label{eq:global-map-step-1} 
  \begin{align}
    \bar p_{k} & =a_{k}\tilde{p}_{i}+b_{k}\tilde{p}_{j}+k_{\perp}\,,\\
    \bar p_{i} & =(1-a_{k})\tilde{p}_{i}\,,\\
    \bar p_{j} & =(1-b_{k})\tilde{p}_{j}\,.
  \end{align}
\end{subequations}
followed by a readjustment involving all particles so as to conserve
momentum~\cite{DSsupplement}, \S\ref{sec:updated-recoil}.
For the original PanGlobal NLL shower, the splitting probability was
given by
\begin{multline}
  \label{eq:split-prob}
  \frac{d\mathcal{P}_{n\to n+1}}{d\ln v } =\!\!\!
  \sum_{\{\itilde,\jtilde\}\in \text{dip}}
  \int d\bar\eta \frac{d\phi}{2\pi}
  \frac{\as(\kT)}{\pi}
  \left(1+ \frac{\as(\kT)\KCMW}{2\pi}\right)
  \\ \times
  \left[f(\bar\eta) a_k P_{\itilde \to ik}(a_k) +
    f(-\bar\eta) b_k P_{\jtilde \to jk}(b_k)
    \right]\,.
\end{multline}
Here $P_{\itilde \to ik}(a_k)$ is a leading-order QCD splitting
function, $\bar \eta = \frac12 \ln a_k/b_k + \text{const.}$, with the
constant arranged so that $\bar \eta=0$ when the emission bisects the
dipole in the event centre-of-mass frame, and
$f(\bar \eta) = 1/(1+e^{-2\bar \eta})$ is a partitioning function.
Additionally, the $\overline{\text{MS}}$ coupling, $\as(\kT)$, uses at
least 2-loop running, and
$\KCMW =\left(67/18-\pi^2/6\right)C_A
  -5/9\,n_f$~\cite{Catani:1990rr}.

\begin{figure}
  \centering
  \includegraphics[width=\columnwidth]{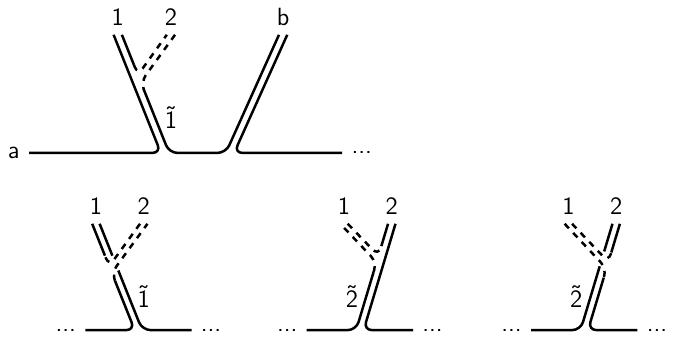}
  \caption{Top: one shower history that produced a proximate $\{1,2\}$
    soft pair.
    Bottom: other histories that could have led to the
    same configuration of momenta, also taken into account in
    correcting the branching.
    The dashed parton is emitted second in the showering history.}
  \label{fig:histories}
\end{figure}

In moving towards higher accuracy, the two relevant elements are
the analogues of the real and virtual corrections in a fixed-order
calculation. 
We focus first on the real term, where we require the shower to
generate the correct double-soft matrix element when two particles are
produced at commensurate angles and (small) energies, while well-separated
from all other particles.

Our approach is illustrated in Fig.~\ref{fig:histories}. Consider the
case where a dipole $ab$ first emits a soft gluon $\tilde{1}$, followed
by a splitting of the dipole $\tilde{1}b$ whereby a new particle $2$
is emitted, and $\tilde{1}$ becomes $1$ after recoil.
When the branching from Eq.~(\ref{eq:global-map-step-1}) produces a
particle $2$ from the $\tilde{1}b$ dipole, if $p_1\ourdot p_2 < p_2\ourdot p_b$, we
select the $\{1,2\}$ pair as the one whose double-soft effective
matrix element needs correcting.
To evaluate the double-soft correction to this configuration, we first
identify all shower histories that could have produced the same nearby
$\{1,2\}$ pair. This includes the history actually followed by the
shower, as well as the case where $2$ was emitted from the $a\tilde{1}$
dipole, and two extra configurations where the shower produced a
particle $\tilde{2}$ before $1$, i.e.\ where, in the second splitting,
gluon $1$ was radiated with $2$ taking the recoil.

Each history $h$ is associated with an effective squared shower matrix
element $|M_{\text{shower},h}|^2$, reflecting the
probability that the shower, starting from the $ab$ system, would
produce the $\{1,2\}$ pair in that order and colour configuration (we
address the question of the flavour configuration below).
$|M_{\text{shower},h}|^2$ is evaluated in the double-soft
limit (\cite{DSsupplement}, \S\ref{sec:PS-effective-ME}).
In principle, emission $2$ should be accepted with probability
\begin{equation}
  \label{eq:acceptance-correction}
  P_\text{accept} = \frac{|M_\ds|^2}{\sum_h
    |M_{\text{shower},h}|^2}. 
\end{equation}
where $|M_\text{\textsc{ds}}|^2$ is the known double-soft matrix
element for emitting the $\{1,2\}$ soft pair from the $ab$
dipole~\cite{Dokshitzer:1992ip,Campbell:1997hg,Catani:1999ss}.
In practice, however, there are regions where the shower
underestimates the true matrix element, leading to
$P_\text{accept} > 1$.
Nevertheless, we find that $P_\text{accept}$ always remains smaller
than some finite value $\Omega$.
We therefore enhance the splitting probability
Eq.~(\ref{eq:split-prob}) by an overhead factor $\Omega$, and
accept the emission with probability $P_\text{accept}/\Omega$.

The numerator and denominator in Eq.~(\ref{eq:acceptance-correction})
are evaluated in the same double-soft limit, defined by rescaling 
$p_1\rightarrow \lambda p_1$, $p_2\rightarrow \lambda p_2$ and taking
the limit $\lambda \rightarrow 0$.
This ensures that $P_\text{accept} = 1$ when $1$ and $2$ are well
separated, thus not affecting regions where the shower was already
correct.

The acceptance procedure is sufficient to ensure the proper generation
of the $\{1,2\}$ kinematics, but not the relative weights of the
$a12b$ and $a21b$ colour connections, which is crucial to reproduce
the pattern of subsequent much softer radiation
from the $\{a,1,2,b\}$ system, as required for NSL accuracy.
To address this problem, we evaluate $F_\shower^{(12)}$, the fraction
of the shower effective double-soft matrix element associated with the
$a12b$ colour connection, and similarly $F_\ds^{(12)}$ for the full
double-soft matrix element, in its large-\nc
limit~\cite{Campbell:1997hg,Catani:1999ss}.
If the shower has generated the $a12b$ colour connection and
$F_\shower^{(12)} > F_\ds^{(12)}$, then we swap the colour
connection with probability 
\begin{equation}
  \label{eq:swap}
  P_\text{swap} = \frac{F_\shower^{(12)} -
    F_\ds^{(12)}}{F_\shower^{(12)}}.
\end{equation}
We apply a similar procedure when the shower generates the $a21b$
connection.
In practice, we precede the colour swap with an analogous procedure
for adjusting the relative weights of $gg$ and $q\bar q$ flavours for
the $\{1,2\}$ pair.
An alternative would have been to apply $P_\text{accept}$ separately
for each colour ordering and flavour combination, however when we
investigated that option for the PanGlobal class of showers, we
encountered regions of phase space where the acceptance probability
was unbounded.
Illustrative plots of the shower matrix element and corrections are given in
the supplemental material~\cite{DSsupplement}, \S\ref{sec:accept-prob}.

Next, we address the question of virtual corrections. 
When $\tilde 1$ is produced in the deep soft-collinear region of the
$ab$ dipole, i.e.\ $\theta_{a\tilde1} \ll \theta_{ab}$ or
$\theta_{\tilde 1b} \ll \theta_{ab}$, the inclusion of $\KCMW$ in
Eq.~(\ref{eq:split-prob}) already accounts for second order
contributions to the branching probability in the soft-collinear
region, as required for NLL accuracy for global event shapes.
However, in general, \KCMW alone is not sufficient when
$\theta_{a1} \sim \theta_{1b} \sim \theta_{ab}$, notably because of
the non-trivial $\etabar$ dependence in Eq.~(\ref{eq:split-prob}) and
the way in which it connects with the overall event momentum $Q$.
Therefore we need to generalise $\KCMW \to K(\Phi_{\tilde 1,ab})$,
where the full $K$ is a function of the kinematics of $\tilde 1$ and
of the opening angle of the $ab$ dipole.
In the same vein as the MC@NLO~\cite{Frixione:2002ik} and
POWHEG~\cite{Nason:2004rx,Frixione:2007vw} methods and their
MINLO~\cite{Hamilton:2012np,Hamilton:2012rf} 
extension, the correct next-to-leading-order (NLO) normalisation
for the emission is given by
\begin{equation}
  \label{eq:K-general}
  K(\Phi_{\tilde 1,ab})
  = V(\Phi_{\tilde 1,ab}) +
  \int d\Phi^\ps_{12/\tilde 1} |M^{\PS}_{12/\tilde 1}|^2
  -
  \Delta^{(\text{\textsc{ps}},1)}_{\tilde 1}.
\end{equation}
Here, $V$ is the exact QCD 1-loop contribution for a single soft
emission, renormalised at scale $\mu = k_{t,\tilde 1}$;
$d\Phi^\ps_{12/\tilde 1} |M^{\PS}_{12/\tilde 1}|^2$ is the
product of shower phase space and matrix element associated with real
$\tilde{1}\to 12$ branching, including double-soft corrections;
and $\Delta^{(\text{\textsc{ps}},1)}_{\tilde 1}$ is the coefficient of
$\as /(2\pi)$ in the fixed-order expansion of the shower Sudakov
factor.
To aid in the evaluation of $K(\Phi_{\tilde 1,ab})$ we make use of two
main elements: firstly, in the soft-collinear limit,
$K(\Phi_{\tilde 1,ab}) \to \KCMW$;
secondly, both $V(\Phi_{\tilde 1,ab})$ and
$\Delta^{(\text{\textsc{ps}},1)}_{\tilde 1}$ are independent of the
rapidity of $\tilde 1$, as long as $\tilde 1$ is soft and (for
$\Delta^{(\text{\textsc{ps}},1)}_{\tilde 1}$) kept at some fixed value
of the evolution scale.
We can therefore reformulate Eq.~(\ref{eq:K-general}) as $K = \KCMW +
\Delta K$, with
\begin{equation}
  \label{eq:DeltaK}
  \Delta K = \int_r
  d\Phi^\PS_{12/\tilde 1} \,|M^{\PS}_{12/\tilde 1}|^2
  - \int_{r_\text{sc}} d\Phi^\PS_{12/\tilde 1_\text{sc}} |M^{\PS}_{1 2/\tilde 1_\text{sc}}|^2.
\end{equation}
In the second term, $\tilde 1_\text{sc}$ is at the same
shower scale $v$ as $\tilde 1$, but shifted by a constant in rapidity with respect
to $ab$ so as to be in the soft-collinear region, wherein
$K(\Phi_{{\tilde 1}_\text{sc},ab}) \rightarrow \KCMW$.
The labels $r$ and $r_\text{sc}$ indicate a regularisation of the
phase space, which should be equivalent between the two terms.
Specifically, we separate $M_\text{DS}$ in
Eq.~(\ref{eq:acceptance-correction}) into correlated and uncorrelated
parts, respectively those involving $C_F C_A$ versus $C_F^2$ colour
factors for the $\bar q gg q$ matrix element.
For the correlated part, we cut on the relative transverse momentum of
$1$ and $2$, while for the uncorrelated part, we cut on the transverse
momentum with respect to the $ab$ dipole and impose
$|\Delta y_{12}| < \Delta y_{\max}$.
In practice we tabulate $\Delta K$ as a function of $\theta_{ab}$,
$\etabar_{\tilde 1}$, and $\phi_{\tilde 1}$, though one could also
envisage on-the-fly evaluation.
We incorporate $\Delta K$ in Eq.~(\ref{eq:split-prob}), through a
multiplicative factor
$1 + \tanh[\frac{\as}{2\pi}\,\Delta K(1-a_k)(1-b_k)]$.
This form keeps the correction positive and bounded.
It also leaves the shower unmodified in the hard-collinear region.

We study the above approach with several variants of the PanGlobal
shower.
All have been adapted relative to Ref.~\cite{Dasgupta:2020fwr} with
regards to the precise way in which they restore momentum conservation
after the map of Eq.~(\ref{eq:global-map-step-1}).
This was motivated by the discovery that in higher-order shower
configurations involving three similarly collinear hard particles, the
original recoil prescription could lead to unwanted long-distance kinematic
side effects.
Details are given in the supplemental material~\cite{DSsupplement},
\S\ref{sec:updated-recoil}, and tests were carried out using the
method of Ref.~\cite{Caola:2023wpj}.

We will consider three variants of the PanGlobal shower: two choices
of the ordering variable, $\sim k_t \theta^{\beta}$ with $\beta=0$
(PG$_{\beta=0}$) and $1/2$ (PG$_{\beta=1/2}$), and also a
``split-dipole-frame'' $\beta=0$ variant (PG$^\text{sdf}_{\beta=0}$),
which replaces $f(\pm \bar\eta) \to f(\pm \eta)$ in
Eq.~(\ref{eq:split-prob}), with $\eta = \frac12 \log a_k/b_k$.
The $\eta=0$ transition region  bisects the dipole in its rest
frame rather than the event frame.
This makes the $\tilde 1 \to 12$ branching independent of the
$\tilde 1$ rapidity in the dipole frame, resulting in $\Delta K = 0$.
Illustrative plots of $\Delta K$ and its impact are given in
Ref.~\cite{DSsupplement}, \S\ref{sec:Delta-K}.
For the three shower variants, the overhead factors $\Omega$
associated with Eq.~(\ref{eq:acceptance-correction}) are respectively
taken equal to $3.1$, $20$ and $4$, independently of the dipole
kinematics.

All results, both with and without double-soft corrections,
include NLO $2$-jet matching~\cite{Hamilton:2023dwb}, which is
required for the NNDL/NSL accuracy that we aim for.
Spin correlations~\cite{Karlberg:2021kwr,Hamilton:2021dyz} are turned
off, because we have yet to integrate them with the double-soft
corrections.
The double-soft corrections are implemented at large-$\nc$, in such a
way as to preserve the full-\nc NLL/NDL accuracies obtained in
Ref.~\cite{Hamilton:2020rcu} for global observables and multiplicities.
All events have (positive) unit weight.

\begin{figure}
  \centering
  \includegraphics[width=\columnwidth,page=1]{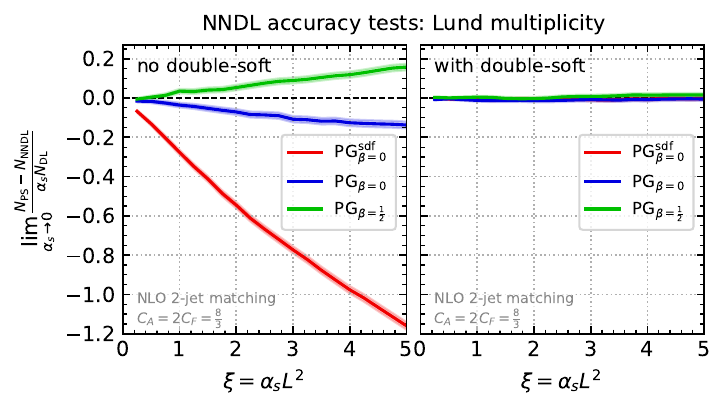}
    \caption{The result of Eq.~(\ref{eq:NNDL-test}) for three variants
    of the PanGlobal shower without double-soft corrections (left) and
    with them (right). 
    The latter are consistent with NNDL accuracy.
    The bands represent statistical errors in an $\as \to 0$
    extrapolation based on four finite $\as$ values.
  }
  \label{fig:nndl-multiplicity}
\end{figure}

To test the enhanced logarithmic accuracy of the shower, the first
observable that we consider is the Lund subjet
multiplicity~\cite{Medves:2022ccw} in $e^+e^- \to q\bar q$ events.
This is a perturbatively calculable observable that is conceptually
close to the experimentally important total charged-particle
multiplicity.
For a centre-of-mass energy $Q$ and a transverse momentum
cutoff $k_t$, the subjet multiplicity has a double-logarithmic
resummation structure $\as^n L^{2n}$ with $L = \ln k_t/Q$.
The PanGlobal showers already reproduce terms up to NDL
$\as^n L^{2n-1}$.
The addition of the double-soft corrections and
matching~\cite{Hamilton:2023dwb} is expected to bring NNDL accuracy,
$\as^n L^{2n-2}$.
To test this, in Fig.~\ref{fig:nndl-multiplicity},
we examine
\begin{equation}
  \label{eq:NNDL-test}
  \lim_{\as \to 0}
  \left. \frac{N_\ps - N_\nndl}{\as N_\dl} \right|_{\text{fixed $\as L^2$}},
\end{equation}
where $N_\ps$ is the parton-shower result and $N_\nndl$ ($N_\dl$) is
the known analytic NNDL (DL) result~\cite{Medves:2022ccw}.
The $\alpha_s\to 0$ limit follows the procedure from earlier work~\cite{Dasgupta:2020fwr}.
Eq.~(\ref{eq:NNDL-test}) is expected to be zero if the parton shower
is NNDL accurate.
The original showers, without double-soft corrections (left),
clearly differ from each other and from zero, by up to $\sim 100\%$.
With double-soft corrections turned on (right), all three
PanGlobal variants are consistent with zero, i.e.\ with NNDL accuracy,
to within $\sim 1\%$.

\begin{figure}
  \centering
  \includegraphics[width=\columnwidth]{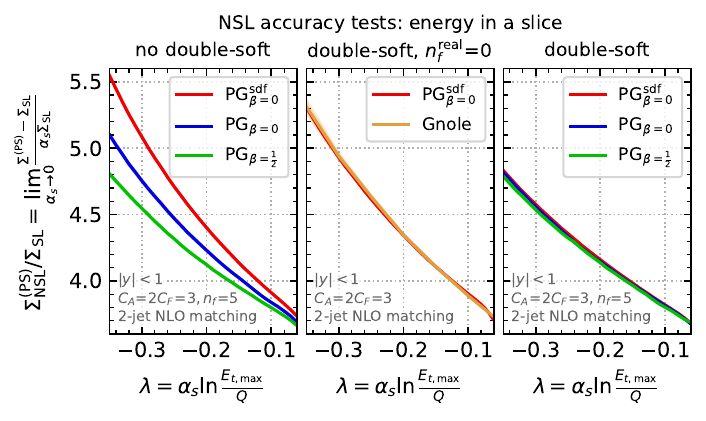}
  \caption{
    Determinations of $\Sigma_\nsl^{(\ps)}/\Sigma_\SL$ for the transverse energy in a slice.
    Left: parton showers without
    double-soft corrections illustrating NSL differences
    between them.
    Middle: with double-soft corrections but $n_f^\text{real}\!=0$ (cf.\ text for details), for
    comparison with the Gnole NSL code.
    Right: with full double-soft corrections, showing NSL agreement between the three
    PanGlobal showers.
  }
  \label{fig:nsl-non-global}
\end{figure}

Next we turn to the study of non-global logarithms at leading
colour.
These were recently calculated at NSL
accuracy~\cite{Banfi:2021owj,Banfi:2021xzn,Becher:2023vrh},
$\as^n L^{n-1}$, and are available in the corresponding ``Gnole''
code~\cite{Banfi:2021xzn}.
We again consider $e^+e^-$ events, and sum the transverse energies ($E_t$) of particles
with $|y| < 1$, where $y$ is the rapidity with respect to an axis
determined by clustering the event to two jets with the Cambridge
algorithm~\cite{Dokshitzer:1997in}.
The fraction of events where the sum is below some $E_{t,\max}$ is
denoted by $\Sigma$ and for a given shower we define
\begin{equation}
  \label{eq:nsl-log-test}
  \Sigma_\nsl^{(\ps)} = 
  \lim_{\as \to 0}
  \left. \frac{\Sigma^{(\ps)} - \Sigma_\SL}{\as}
  \right|_{\text{fixed $\as L$}}\hspace{-3.5em},\hspace{2.8em}
  \quad L \equiv \ln \frac{E_{t,\max}}{Q}.
\end{equation}
Fig.~\ref{fig:nsl-non-global} (left) shows $\Sigma_\nsl^{(\ps)}/\Sigma_\SL$ for
our three PanGlobal variants without double-soft corrections.
As expected, they differ.

Fig.~\ref{fig:nsl-non-global} (middle) compares our
PG$_{\beta=0}^\text{sdf}$ shower with double-soft corrections to the
NSL Gnole code, showing good agreement, within $<1\%$.
Gnole has $n_f=0$ in the real contribution and counterterm, but
keeps the full $n_f=5$ in the running of the coupling and inclusive
$\KCMW$ (``$n_f^\text{real}=0$'').
Among our showers it is relatively straightforward to make the same
choice with PG$_{\beta=0}^\text{sdf}$, in particular because
$\Delta K = 0$.
Also, Gnole uses the thrust axis, while we use the jet axis;
this is beyond NSL as the two axes coincide for hard three-parton
events.
\logbook{}{see
  2020-eeshower/analyses/rapidity-slice-double-soft-nll-tests/axis-check/axis-check.pdf
and associated log file (which prints out largest difference)}

Fig.~\ref{fig:nsl-non-global} (right) shows the results from our three PanGlobal showers
with complete (full-$n_f$) double-soft corrections included.
They agree with each other to within 1\% of the NSL contribution, providing a powerful
test of the consistency of the full combination of the double-soft
matrix element and $\Delta K$ across the variants.
That plot also provides the first NSL calculation of non-global
logarithms to include the full $n_f$ dependence.
An extended selection of results and comparisons is provided in
\S\ref{sec:ref-NGL-NSL} of Ref.~\cite{DSsupplement}.

\begin{figure}[t]
  \centering
  \includegraphics[width=\columnwidth,page=1]{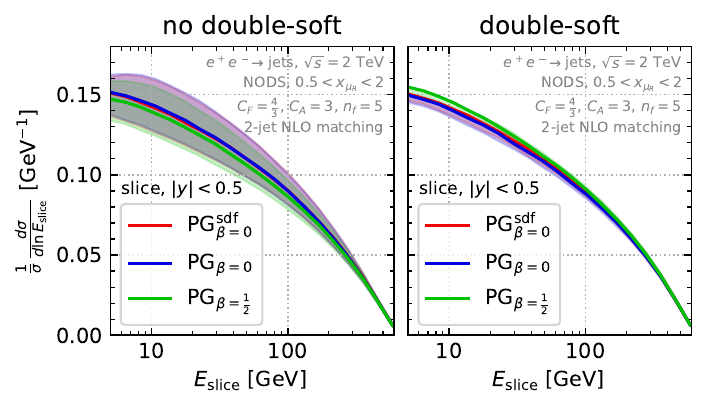}
  \caption{Distribution of energy in a slice $|y|<0.5$ for the
    PanGlobal shower without double-soft corrections (left) and with
    them (right).
    The bands represent renormalisation scale variation, with NLO
    scale-compensation enabled only for the results with double-soft
    corrections.
  }
  \label{fig:nsl-non-global-pheno}
\end{figure}

We close with a brief examination of the phenomenological implications
of the advances presented here.
We consider $e^+e^- \to Z^* \to \text{jets}$ at $Q = 2 \TeV$.
This choice is intended to help gauge the size of non-global effects
at the energies being probed today at the LHC.
Fig.~\ref{fig:nsl-non-global-pheno} shows results for the distribution
of energy flow in a rapidity slice, defined with respect to the 2-jet
axis, without double-soft corrections (left) and with them, i.e.\ at
NSL accuracy (right).
It uses the nested ordered double-soft (NODS) colour scheme, which while not full-$\nc$ accurate
for non-global logarithms, numerically coincides with the full-$\nc$
single logarithmic results of
Refs.~\cite{Hatta:2020wre,Hagiwara:2015bia,Hatta:2013iba}, to within
their percent-level numerical accuracy~\cite{Hamilton:2020rcu}.
With a central scale choice (solid lines), the impact of the NSL
corrections is modest.
This is consistent with the observation from
Fig.~\ref{fig:nsl-non-global} that the NLL PanGlobal showers are
numerically not so far from NSL accurate.
However, the NSL double-soft corrections do bring a substantial
reduction in the renormalisation scale uncertainty, from about $10\%$
to just a few percent.
Conclusions are similar for $H^* \to gg$.

The results here provide the first demonstration that it is possible
to augment parton-shower accuracy beyond NDL/NLL.
Specifically, our inclusion of real and virtual double-soft effects
has simultaneously brought NNDL/NSL accuracy for two
phenomenologically important classes of observable: multiplicities,
and energy flows as relevant for isolation.
It has also enabled the first leading-colour, full-$n_f$ predictions
for NSL non-global logarithms.
Overall, our methods and results represent a significant step towards
a broader future goal of general next-to-next-to-leading logarithmic
accuracy in parton showers.

\begin{acknowledgments}
  We are grateful to Pier Monni for the very considerable work
  involved in the comparisons of NSL non-global logarithms for the
  energy flow in a slice, in particular with regards to the extraction
  of pure NSL results from the Gnole code.
  We are grateful to our PanScales collaborators (Melissa van
  Beekveld, Mrinal Dasgupta, Fr\'ed\'eric Dreyer, Basem El-Menoufi,
  Jack Helliwell, Rok Medves, Pier Monni, Alba Soto Ontoso and Rob
  Verheyen), for their work on the code, the underlying philosophy of
  the approach, the adaptations of the PanGlobal shower discussed in
  Ref.~\cite{DSsupplement}, \S\ref{sec:updated-recoil} and comments on
  this manuscript.
  This work has been funded by the European Research Council (ERC)
  under the European Union's Horizon 2020 research and innovation
  program (grant agreement No 788223, SFR, KH, AK, GPS, GS, LS),
  by a Royal Society Research Professorship
  (RP$\backslash$R1$\backslash$180112, GPS and LS) and by the Science
  and Technology Facilities Council (STFC) under grant ST/T000864/1
  (GPS) and ST/T000856/1 (KH).
  SFR, KH, GPS and LS benefited from the hospitality of the Munich
  Institute for Astro-, Particle and BioPhysics (MIAPbP) which is
  funded by the Deutsche Forschungsgemeinschaft (DFG, German Research
  Foundation) under Germany's Excellence Strategy -- EXC-2094 --
  390783311.
  KH, GPS and LS also wish to thank CERN's Theoretical Physics
  Department for hospitality while the work was being completed.
\end{acknowledgments}

\bibliographystyle{apsrev4-2}
\bibliography{MC}

\input{supplementary_material}

\end{document}

%% file: supplementary_material.tex
\newpage

\onecolumngrid
\newpage
\appendix

\makeatletter
\renewcommand\@biblabel[1]{[#1S]}
\makeatother

\section*{Supplemental material}

\subsection{Event-wide momentum conservation for the PanGlobal shower}
\label{sec:updated-recoil}

Compared to the original formulation in Ref.~\cite{Dasgupta:2020fwr},
this study introduces a revised rescaling procedure for the kinematic
map of the PanGlobal parton shower.
Here we outline an important flaw identified in the original PanGlobal
map, before detailing two amended versions which rectify it, one of
which has been used for all results in this article.

\subsubsection{Unsafety of the original formulation}\label{sec:earlier-unsafe}

Given a set of splitting variables $\{v,\bar\eta,\phi\}$ generated
according to Eq.~(\ref{eq:split-prob}), the original PanGlobal
kinematic map of Ref.~\cite{Dasgupta:2020fwr} proceeds by first
constructing a set of intermediate post-branching momenta according
to Eq.~(\ref{eq:global-map-step-1}), conserving only longitudinal
momentum within the emitting dipole system. 
All momenta in the event are subsequently rescaled by a factor
$ r = \sqrt{Q^2/(Q + k_{\perp})^2}$,
which serves to bring the total invariant mass of the event back to
$Q^2$.
Finally, a Lorentz boost is applied to all particles restoring the
total four-momentum in the event, from $r ( Q+k_\perp )$ to $Q$.

In all kinematic limits where shower emissions are well-separated in
logarithmic phase-space, i.e.\ all limits pertaining to NLL accuracy,
the rescaling factor, $r$, tends to one. 
While investigating kinematic configurations relevant beyond NLL
accuracy, we found a specific sequence of emissions, involving three
similarly-collinear hard particles, that leads to $r$ differing from
one by an $\mathcal{O}(1)$ amount.

The fact that a double-unresolved configuration produces long-range
(event-wide) $\mathcal{O}(1)$ momentum shifts is in contradiction with
the fundamental construction principle of the PanScales showers.
This spurious effect evades NLL-accuracy tests, which are by definition
insensitive to triple-collinear corrections.
Nevertheless, since configurations with three similarly-collinear hard
particles can be generated at arbitrarily small angles, the resulting large
rescalings technically mean that the original PanGlobal rescaling
prescription violates infrared-and-collinear safety.

The origin of large rescaling coefficients can be simply illustrated
as follows.
Consider the situation where, in the event frame, one has a
highly-energetic dipole $\itilde \jtilde$, with $E_\itilde\sim Q$,
$E_\jtilde\sim Q$, and an opening angle
$\theta_{\itilde\jtilde}\ll 1$.
We then radiate a hard-collinear emission $k$ from the $\itilde
\jtilde$ dipole using the kinematic map of
Eq.~(\ref{eq:global-map-step-1}).
The transverse momentum of the emission (with respect to the dipole)
is given by $k_t^2=a_kb_k m_{\itilde\jtilde}^2$ which, in the
commensurate triple-collinear limit ($a_k \sim b_k \sim 1$), becomes
$k_t^2\sim m_{\itilde \jtilde}^2 \sim \theta_{\itilde \jtilde}^2Q^2 $.
Consequently, in the $\itilde\jtilde$ dipole rest frame, the original
dipole momenta and the transverse contribution are all commensurate,
and of order $\theta_{\itilde \jtilde} Q$.
Boosting to the event frame, all momenta receive a large
Lorentz boost factor
$\gamma\sim (E_\itilde+E_\jtilde)/m_{\itilde\jtilde}\sim
1/\theta_{\itilde \jtilde}$.
In particular, this results in the transverse component having an
energy of order $Q$ in the event frame, or equivalently
$k_\perp\ourdot Q\sim (1/\theta_{\itilde \jtilde})(\theta_{\itilde
  \jtilde}Q)\ourdot Q \sim Q^2$.
Since transverse components are not conserved by the map of
Eq.~(\ref{eq:global-map-step-1}), one obtains a rescaling factor $r$
that significantly departs from one.

Below, we introduce two small adaptations of
the original kinematic rescaling that each cure this issue.

\subsubsection{Updated prescriptions with local rescaling}\label{sec:new-rescaling-schemes}

The essential idea of the updated prescriptions that we introduce below is
to avoid spuriously large rescalings of the event as a whole by
absorbing the potentially large energy component of the transverse
momentum $k_\perp$ via a dipole-local rescaling.
We start with the PanGlobal momentum map from
Eq.~(\ref{eq:global-map-step-1}).

In the first of our two updated prescriptions, instead of applying a
global rescaling to all the momenta in the event, we rescale
$\bar p_i$, $\bar p_j$ and $\bar p_k$ by a local rescaling factor
$\rescl$.
As with the original variant, we fix the rescaling by imposing that
the invariant mass of the event is returned to $Q^{2}$.
Defining time-like four-momenta
$\bar{p}_{ijk}=\bar{p}_{i}+\bar{p}_{j}+\bar{p}_{k}$ and
$\tilde{p}_{m}=Q-\tilde{p}_{i}-\tilde{p}_{j}$, we find
\begin{equation}
  \rescl = 
  \frac{ -\tilde{p}_{m}\ourdot\bar{p}_{ijk}
         + \sqrt{ (\tilde{p}_{m}\ourdot\bar{p}_{ijk})^{2} +
                  \bar{p}_{ijk}^{2} \, (Q^{2}-\tilde{p}_{m}^{2})
                }
       }
       {\bar{p}_{ijk}^{2}}.
  \label{eq:PG-variants-r-factor}
\end{equation}
After the (local) rescaling the whole event undergoes a Lorentz
boost so as to restore the total four-momentum of the event to its
original value $Q$ (as in the original PanGlobal prescription).
This is the prescription used throughout this paper.

A second possibility is as follows.
Noting that the origin of the issue with global rescaling
lies in the large energy component of $k_\perp$, one can proceed with
a hybrid approach whereby one first applies a local rescaling
$\rescl'$ to restore the dipole energy in the original event frame, i.e.\
imposing $\rescl'\, \bar p_{ijk}\ourdot Q = (\tilde p_i+\tilde p_j)\ourdot
Q$.
One subsequently applies a global rescaling $\rescg'$ to restore
the invariant mass $Q^2$ of the whole event, followed by a Lorentz
boost, as in the original formulation.
We find
\begin{equation}
  \rescl' = \frac{(\tilde p_i+\tilde p_j)\ourdot Q}{ \bar p_{ijk}\ourdot Q},
  \qquad\text{ and }\qquad
  \rescg' = \sqrt{\frac{Q^2}{(\tilde{p}_{m}+\rescl' \bar{p}_{ijk})^2}}.
  \label{eq:PG-variants-rprime-factors}
\end{equation}
Note that similar rescaling strategies can be adopted in the
formulation of the PanGlobal shower for $pp$ collisions,
deep-inelastic scattering and
vector-boson fusion (see Ref.~\cite{vanBeekveld:2023lfu} for details).

The specific choice between the two updated schemes does not affect
the tests carried out in this paper.
This owes to the following two facts:
firstly, in the double-soft region all rescalings go to one (both in
the original formulation and in the two amended versions introduced
here);
secondly, for the special case of the first emission, where there are
only three post-branching momenta, all three maps are identical,
meaning that the matching procedure is also
unaffected~\cite{Hamilton:2023dwb}.

To validate the new rescaling prescription, we have performed
numerical tests similar to the infrared-and-collinear-safety
tests carried out for flavoured-jet algorithms in
Ref.~\cite{Caola:2023wpj}. While the original prescription fails these
tests, the new approaches described here pass them.
Additionally, we have extended those tests to automate
the verification of the more stringent PanScales criteria, proceeding
as follows.
First, we generate an initial set of emissions. Second, we add new
emissions widely separated from the initial ones by at least a
distance $\Delta$ in a logarithmic phase-space (equivalently, the Lund
plane).
We then check that the recoil induced on the initial emissions
decays exponentially with increasing $\Delta$.
While the original global rescaling approach fails these tests (as do
standard dipole showers), the new variants pass.
Note that if we further impose that the additional emissions are
themselves widely separated in logarithmic phase-space, both the
original and new rescaling prescriptions pass the test, in accordance
with the expectations from the NLL matrix-element tests performed in
earlier work.

%======================================================================
\subsection{Matrix element tests and $\Delta K$}
\label{sec:accept-prob-and-delta-K}

This section records expressions for the effective shower matrix
elements, before giving illustrative results regarding the validation
of the two key elements of our double-soft corrections, namely the
real matrix-element corrections and the virtual $\Delta K$ correction. 

%----------------------------------------------------------------------
\subsubsection{Effective shower matrix elements}

\label{sec:PS-effective-ME}

The total effective squared shower matrix element in
Eq.~(\ref{eq:acceptance-correction}), at large-\nc, describing the
radiation of a double-soft gluon or quark pair, $\left\{ 1,2\right\} $,
from a dipole $ab$, is most conveniently expressed as a sum of
contributions associated to the two contributing colour connections,
$a12b$ and $a21b$\,: 
\begin{equation}
   \sum_{h}\,|M_{\mathrm{shower},h}|^{2} = 
   \sum_{h\in a12b}
   |M_{\mathrm{shower},h}^{(12)}|^{2}
   +
   \sum_{h\in a21b}
   |M_{\mathrm{shower},h}^{(21)}|^{2}.
   \label{eq:PS-effective-ME-total}
\end{equation}
The contribution associated to the $a21b$ colour ordering has the
form 
\begin{align}
   \sum_{h\in a21b}
   |M_{\mathrm{shower},h}^{(21)}|^{2} = &
   (8\pi\alpha_{s})^{2} \,
   2C_{1} \,
   \mathcal{J}
   \left[f(\hat\eta_{2})\,2C_{2}\,\delta_{a\rightarrow a2} + 
         f(-\hat\eta_{2})\,z_{2}\,P_{\tilde{1}\rightarrow12}(z_{2})
   \right]
   \Theta(v_{\tilde{1}}>v_{2})
   \, +\, 
   \{1\leftrightarrow2\} \, ,
   \label{eq:PS-effective-ME-a21b}
\end{align}
wherein $\mathcal{J}$, $\bar{\eta}_{2}$, and $z_{2}$ are given
in terms of invariants $s_{ij}=2p_{i}\ourdot p_{j}$ and $s_{i}=2p_{i}\ourdot Q$
as follows
\begin{align}
   \mathcal{J} = &
   \frac{s_{ab}}{s_{b1} (s_{a1}+s_{a2})} 
   \frac{s_{a1}}{s_{a2} s_{12}}, & 
   \bar\eta_{2} = &
   \frac{1}{2}\log\left(\frac{s_{a}s_{12}}{s_{1}s_{a2}}\right)
   = \eta_2 + \frac{1}{2}\log\left(\frac{s_{a}s_{a1}}{s_{1}(s_{a1}+s_{a2})}\right), &
   z_{2} = &
   \frac{s_{a2}}{s_{a1}+s_{a2}} .
   \label{eq:PS-effective-ME-a21b-J-etabar2-z2}
\end{align}
The angular dipole partitioning functions, $f(\pm\bar{\eta}_{2})$,
and the splitting functions, $P_{\tilde{1}\rightarrow12}(z_{2})$,
are as given in Ref.~\cite{Dasgupta:2020fwr}; for PG$_{\beta=0}$
and PG$_{\beta=1/2}$ we have $\hat\eta_{2} = \bar\eta_2$, as in
Eq.~(\ref{eq:split-prob}), while for PG$^\text{sdf}_{\beta=0}$
this is modified to $\hat\eta_{2} = \eta_2$.
The $\delta_{a\rightarrow a2}$ factor is equal to one when the radiated
pair comprises of gluons, and zero when it consists of quarks. 
Colour factors $C_1$ and $C_2$ are both equal to $C_{F}$, with 
$C_{A}=2C_{F}$ in the large-\nc limit.

The final theta function in Eq.~(\ref{eq:PS-effective-ME-a21b})
reflects the ordering of the emissions with respect to
the shower evolution variable $v$. The evolution variable was defined
in terms of $k_{t}$ and $\bar\eta$ in Ref.~\cite{Dasgupta:2020fwr}
through $k_{t}=\rho ve^{\beta|\bar{\eta}|}$.
$\Theta(v_{\tilde{1}}>v_{2})$ in Eq.~(\ref{eq:PS-effective-ME-a21b}) is then
fully specified given Eq.~(\ref{eq:PS-effective-ME-a21b-J-etabar2-z2})
for $\bar{\eta}_{2}$ together with the following additional components 
\begin{align}
   k_{t,\tilde{1}}^{2}
   &= \frac{s_{b1}(s_{a1}+s_{a2})^{2}}{s_{ab}s_{a1}}, & 
   \rho_{\tilde{1}}
   &= \left( \frac{s_{a}s_{b}}{Q^{2}s_{ab}} \right)^{\frac{\beta}{2}}, &
   \bar{\eta}_{\tilde{1}}
   &=\frac{1}{2}\log\left(\frac{s_{a}s_{b1}}{s_{b}s_{a1}} \right),
   &k_{t,2}^{2}
   &= \frac{s_{a2}s_{12}}{s_{a1}}, &
   \rho_{2}
   &= \left(\frac{s_{a}s_{1}}{Q^{2}s_{a1}}\right)^{\frac{\beta}{2}}.
   \label{eq:PS-effective-ME-kt2-rho2}
\end{align}

Finally, the contribution to the total effective squared
shower matrix element arising from the $a12b$ colour
ordering, $\sum_{h\in a12b}\,|M_{\mathrm{shower},h}^{(12)}|^{2}$,
is given by swapping $a\leftrightarrow b$ everywhere
on the right-hand side of Eq.~(\ref{eq:PS-effective-ME-a21b}).

%----------------------------------------------------------------------
\subsubsection{Real double-soft matrix-element corrections}

\label{sec:accept-prob}

\begin{figure}
  \includegraphics[width=0.446\textwidth,page=1]{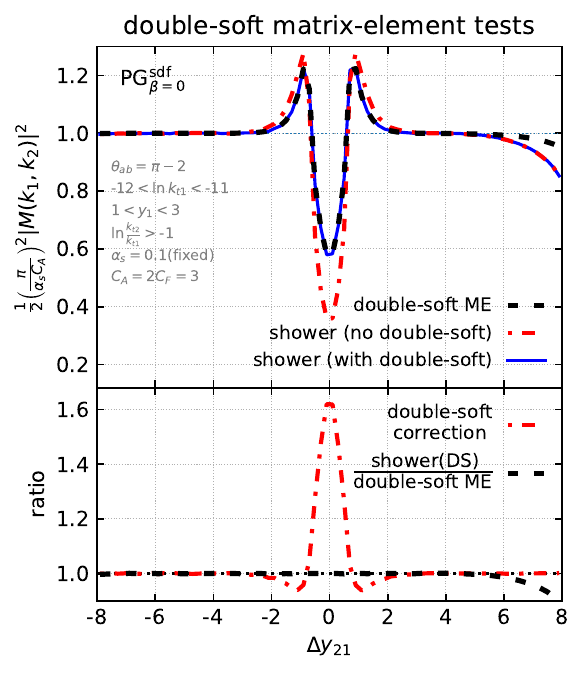}\hfill
  \includegraphics[width=0.525\textwidth,page=2]{figures/me-tests.pdf}
  \caption{Left: rate of emission before (dashed-dot red) and after
    (solid blue) double-soft matrix-element corrections at
    $\mathcal{O}(\alpha_s^2)$. Right: decomposition in terms
    of colour flows and flavour channels.
  }\label{fig:ME-tests}
\end{figure}

In Fig.~\ref{fig:ME-tests} (left) we provide an illustration of one of
the matrix element tests that we have carried out.
It is shown for the PG$^\text{sdf}_{\beta=0}$ shower.
We start with a dipole $ab$ with an opening angle of
$\theta_{ab} = \pi-2$ in the event centre-of-mass
frame.
From that system, we generate two soft emissions, and
select those configurations where the higher-$k_t$ emission ($1$) is in a window $-12
< \ln k_{t1}/Q < -11$ and $1 < y_1 < 3$, while the lower-$k_t$
emission ($2$) satisfies $\ln k_{t2}/k_{t1} > -1$.
We determine transverse momenta and rapidities in the dipole
centre-of-mass frame and, for the purpose of Fig.~\ref{fig:ME-tests},
restrict our attention to configurations for which, in that frame, the
two emissions are both in the $ab$ dipole's primary Lund plane~\cite{Dreyer:2018nbf}.
The upper panel shows the differential distribution of the rapidity
difference between the two emissions, $\Delta y_{21} = y_2 -
y_1$. 
 
The results in Fig.~\ref{fig:ME-tests} have been normalised to
$2\,\!(\as C_A/\pi)^2$ which is the expected result for large
$\Delta y_{21}$, at large-\nc, as long as particle $2$ is still
soft.
The red (dot-dashed) curve shows the default shower, without any
double-soft correction, while the black curve shows the actual
double-soft matrix element.
Both are shown averaged over $\phi_2$, the azimuth of particle $2$.
The shower and exact double-soft matrix element differ for $\Delta
y_{21}$ in the vicinity of zero.
The red dot-dashed curve in the lower panel shows the ratio of the two
curves, illustrating the need for $\order{1}$ corrections at small and
moderate $\Delta y_{12}$ values.
As $|\Delta y_{21}|$ becomes larger, all curves tend to the same limit
corresponding to independent emission.
Once particle 2 starts to become hard, and the physical phase space
boundary is approached, $\Delta y_{21} \, {\scriptstyle \gtrsim} \, 6$, all
three predictions begin to depart from that of the independent emission
picture, with small technical kinematic cuts also playing a role in that region.
Crucially, however, throughout this hard-collinear region the
shower predictions with and without double-soft corrections are seen
to be in perfect agreement.

When the shower is run with the (fully-differential) double-soft
correction factor (upper panel, blue solid line), one sees that it agrees perfectly
with the double-soft matrix element for moderate $\Delta y_{21}$.
One important point is that at large positive $\Delta y_{21}$, the
correction factor does not modify the shower, even though the shower
and the double-soft matrix element differ: in that limit, where the
hard-collinear splitting function corrections are relevant, the
shower already provides the correct answer, and it is important to
maintain that correct answer.

The right-hand plot shows the same differential distribution but
broken into flavour and colour channels.
Let us start with the upper left panel, which shows the $\bar q g_1
g_2 q$ channel, where the particle labelled $1$ is always the one with
larger transverse momentum, and the order of the particles corresponds to
the order of the colour connections.
Of particular interest is the region of negative $\Delta y_{21}$, i.e.\
where the rapidity ordering is opposite to the colour ordering.
In this region the true double-soft matrix element is strongly
suppressed, as one would expect.
However, the shower's suppression is parametrically stronger.
The pattern is similar in the top-right panel for the opposite
$\bar q g_2 g_1 q$ colour ordering at positive $\Delta y_{21}$.
Had we attempted to correct the shower for each colour-channel
separately, there would have been regions where the acceptance
probability in Eq.~(\ref{eq:acceptance-correction}) would have become
arbitrarily large.
Instead the approach of Eq.~(\ref{eq:swap}) ensures that we only have
to make an occasional swap of the colour ordering.
The lower panels show the analogous curves for double-soft quark
production.

%----------------------------------------------------------------------
\subsubsection{$\Delta K$ and evaluation of its impact}
\label{sec:Delta-K}

\begin{figure}
  \includegraphics[width=0.33\textwidth]{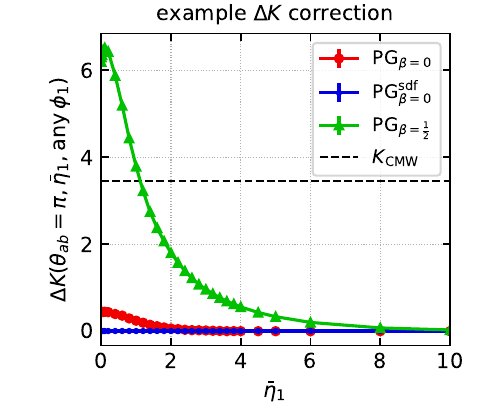}\hfill
  \includegraphics[width=0.33\textwidth,page=1]{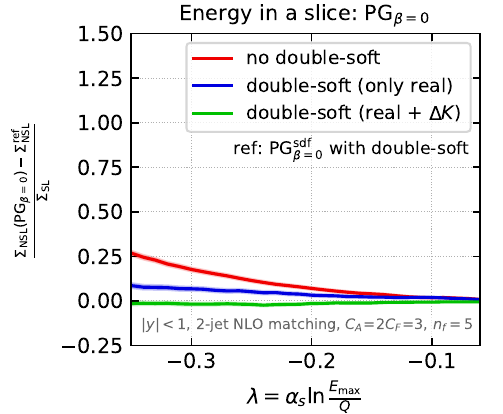}
  \includegraphics[width=0.33\textwidth,page=2]{figures/slice-deltaK-effect.pdf}
  \caption{
    Left: Plot of the NLO $\Delta K$ correction,
    Eq.~(\ref{eq:DeltaK}), for three variants 
    of the PanGlobal shower, as a function of the rapidity
    $\bar\eta_1$ of a soft
    emission from a back-to-back dipole.
    $\KCMW$ is given for reference.
    Centre/right: impact of different parts of the double-soft
    correction on the NSL contributions for the transverse energy in a
    slice, showing the difference between PG$_{\beta=0}$ (centre) or
    PG$_{\beta=\frac12}$ (right) and a reference NSL-accurate shower.
  }\label{fig:DeltaK}
\end{figure}

Recall that for a soft emission probability (from a $\bar q q$ dipole)
as given in Eq.~(\ref{eq:split-prob}), NSL accuracy requires an
extra $(1 + \Delta K \as/2\pi)$ correction factor.
Fig.~\ref{fig:DeltaK} (left) shows the size of the $\Delta K$
contribution, Eq.~(\ref{eq:DeltaK}), for our three PanGlobal shower
variants.
It is plotted as a function of the rapidity, $\bar \eta_1$
of the soft emission, in the case of a back-to-back parent dipole.
The shower with the largest correction is PG$_{\beta=\frac12}$, but
for the configuration shown here, that correction remains relatively
modest, at most a factor of about $(1 + \as)$ for
$\bar \eta_1 = 0$.
The correction for PG$_{\beta=0}$ is much smaller.
The PG$_{\beta=0}^\text{sdf}$ variant has the property that $\Delta K$
is identically zero, a consequence of the fact that the shower's second
emission probability is independent of the rapidity of the first
emission, causing the two terms in Eq.~(\ref{eq:DeltaK}) to exactly
cancel. 

Fig.~\ref{fig:DeltaK} (centre and right) illustrates the separate impact of the
double-soft real matrix element and $\Delta K$ corrections on the
slice observable of Fig.~\ref{fig:nsl-non-global}, for
PG$_{\beta=0}$ (centre) and PG$_{\beta=\frac12}$ (right). 
It shows the difference in NSL contributions between the
PG$_{\beta}$ result and an NSL-accurate reference, which is
taken to be the PG$_{\beta=0}^\text{sdf}$ shower including the full
double-soft corrections.
The red curve shows the difference with no double soft corrections at
all, illustrating e.g.\ the fortuitous near agreement with the full NSL
result for PG$_{\beta=\frac12}$.
Turning on the real double-soft corrections (blue curve) introduces a
highly visible effect, bringing the PG$_{\beta=0}$ result in better
agreement with the full NSL but causing a significant departure from
NSL in the PG$_{\beta=\frac12}$ case.
Including also the $\Delta K$ correction (green curve) results in
agreement with the NSL result for both showers.
The sign of the $\Delta K$ effect is consistent with the left-hand
plot: $\Delta K$ is always positive, 
\logbook{}{Also for other $\eta_\text{skew}$ and $\phi$ values, says Gregory!}%
and the resulting higher emission probability reduces the
value of $\Sigma$.

Finally, let us comment on the numerical accuracy of our results.
For $\lambda=-0.35$, we find $\Sigma_\nsl/\Sigma_\SL=4.832 \pm 0.004$
(PG$_{\beta=0}^\text{sdf}$), $4.817 \pm 0.010$ (PG$_{\beta=0}$) and
$4.787\pm 0.014$ (PG$_{\beta=\frac{1}{2}}$), where the quoted
uncertainties are purely statistical, as obtained from a cubic
polynomial extrapolation $\alpha_s\to 0$.
These numbers are roughly within $2\sigma$ of each other.
Note however that for PG$_{\beta=\frac{1}{2}}$, we found the
convergence with $\alpha_s$ to be slower, making the extraction
numerically more challenging.
Accordingly, one should also keep in mind that this comes with
additional systematic effects.
For example, we observed that varying the set of $\alpha_s$ values
yields variations in $\Sigma_\nsl/\Sigma_\SL$ of the order of
$0.01$. We also estimated the effect of varying $\Delta K$ within its
numerical uncertainty to be of order $0.005$.
In all cases, we see a convincing agreement to within $1\%$ relative
to the size of the NSL correction.

\logbook{}{Here are some numbers for the slice ($\lambda=-0.35$):
PG$_{\beta=0}^\text{sdf}$ gives $4.832 \pm 0.004$, PG$_{\beta=0}$
gives $4.817 \pm 0.010$, i.e.\ within a bit more than $1\sigma$ (even
wo systematics).
Both use $\alpha_s=0.002,0.01,0.016,0.02$ for the extrapolation.
For PG$_{\beta=0.5}$, we get $4.787\pm 0.014$ for
$\alpha_s=0.001,0.004,0.008,0.01$ (our quoted result) $4.794\pm 0.028$
for $\alpha_s=0.001,0.002,0.004,0.008$ $4.792\pm 0.022$ for
$\alpha_s=0.001,0.002,0.008,0.01$.  }

%======================================================================
\subsection{Reference NSL results for non-global logarithms}
\label{sec:ref-NGL-NSL}

\begin{figure}
  \centerline{
  \includegraphics[width=0.8\textwidth, page=1]{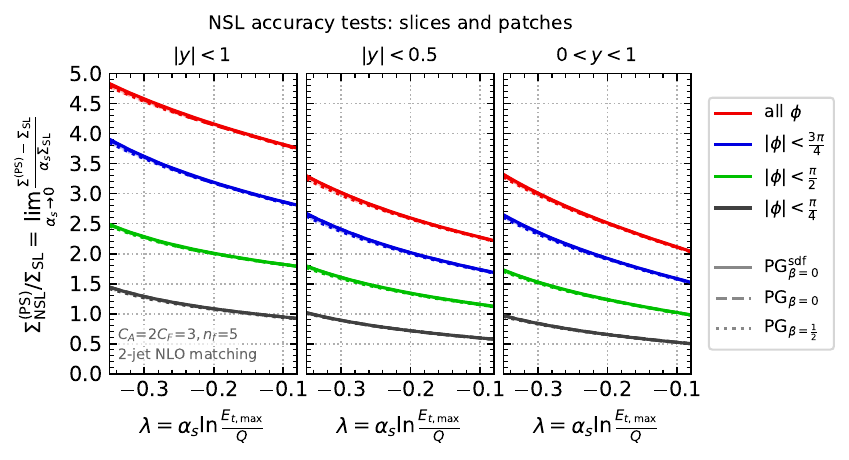}
  }
  \caption{Next-to-single non-global logarithms contributions
  for the transverse energy in patches of given rapidity and azimuthal
  angle range. Each panel corresponds to a fixed rapidity acceptance
  and shows different ranges in azimuthal angle (different colours) as
  well as each of the three PanGlobal showers including double-soft
  corrections (different line styles).}\label{fig:extra-nsl-plots}
\end{figure}

In this last section, we provide additional results for non-global
observables. We consider the transverse energy in a square patch of
fixed extent in rapidity and azimuthal angle. In each case, we study
the next-to-single logarithmic contribution, normalised to the
single-logarithmic result,
$\Sigma_{\text{\textsc{nsl}}}/\Sigma_{\text{\textsc{sl}}}$.
We have extracted $\Sigma_{\text{\textsc{nsl}}}$ using the same
variants of the PanGlobal shower as in the main text.
Our results are presented in Fig.~\ref{fig:extra-nsl-plots}, showing
an excellent degree of agreement, at the 1-2\% level, between the
showers across the whole set of observables.
These can also serve as reference results for future studies of
non-global logarithms at NSL accuracy.

%%% Local Variables:
%%% TeX-master: "panglobal-double-soft"
%%% End:

% LocalWords:  parametrically Eq NLL PanScales NLO NSL